# Grain size determination of superconducting $MgB_2$ powders from magnetization curve, image analysis and surface area measurement

Maurizio Vignolo[1,*], Gianmarco Bovone[2], Emilio Bellingeri[1], Cristina Bernini[1], Gennaro Romano[1,#], Maria Teresa Buscaglia[3], Vincenzo Buscaglia[3] and Antonio Sergio Siri[1,2]

[1] CNR-SPIN, C.so Perrone 24, I-16152 Genova, Italy

[2] University of Genoa, DiFi, Via Dodecaneso 33, I-16146 Genova, Italy

[3] CNR-IENI, Via De Marini 6, I-16149 Genova, Italy

\# now external contractor at ITER Organization, CEO at SUST s.r.l.

\* corresponding author: Tel. +39 010 6598 790, e-mail maurizio.vignolo@spin.cnr.it

**Abstract:** The present article reports a method for the average grain size evaluation of superconducting nano-particles through their magnetic properties. The use of SQUID magnetometry to determine the average $MgB_2$ particle size was investigated and the results compared with those achieved through other techniques. In particular the data obtained from zero field cooled magnetization measurement as function of the temperature were compared with the results obtained by scanning electron microscopy and Brunauer-Emmett-Teller techniques.
The particle magnetization was measured by a commercial SQUID magnetometer in magnetic field (1 mT) and temperatures ranging from 5 to 50 K dispersing the powders in a grease medium. The grain size is obtained by fitting the data taking into account the Ginzburg-Landau temperature dependence of the London penetration depth. Variations on typical modeling parameters were explored in order to gain a better picture of the average grain size and the effectiveness of various measurement techniques.
We find that it is possible to use the magnetization measurements to determine the average grain size even if the SEM image analysis allows extracting more information about the grain size distribution. Furthermore a Matlab routine has been developed in order to get automatic analysis of SEM images.

## 1. Introduction

The $MgB_2$ superconductor exhibits grain-boundary flux pinning [1-3] making it of great interest to qualitatively describe the nano-structural morphology of this material. In general the control of particles grain size and morphology is of crucial importance, especially from the industrial point of view. In fact varying the particles volume from the micrometric to nanometric scale a consequent change in physical properties arise [4]. As consequence of that many performances related grains of engineering materials critically depend on the scale and the morphology of the constituent particles. The observation of changes in the dimension of specific types of grains, as well as the average grain size and also the particle size population are information of vital importance not only from a quality point of view but also to predict the behavior of materials. For example in the milling process it is fundamental to know the final grain size in order to understand how the milling parameters work on the grain dimension. The same is necessary when different synthesis processes are compared in order to establish the best one.

The main driving forces behind our interest are both the development of improved synthesis techniques and an increasing in the number of potential applications for the boron nano-particles of superconducting $MgB_2$. In particular for our research group it is fundamental to establish the average grain size and the particles distribution of the $MgB_2$ powders synthesized in our laboratory following the patented method described in reference [5], and used to produce C-doped $MgB_2$ powders [6]. Furthermore the grain size plays a crucial role in the multifilamentary manufacturing to obtain very fine filaments, as described by Malagoli *et al* and Kováč *et al* in their respective papers [7-9].

There are various methods by which particles grain size distribution may be determined, usual experimental techniques include direct imaging analysis [10, 11], bulk scattering techniques [10], surface area determination [12] and macroscopic magnetic measurements [13]. Transmission electron microscopy (TEM) analysis offers a clear visual evidence of the particles grain size but only a limited number of particles per sample can be analyzed. It is often the case that a heterogeneous sample compromises the observed distribution; so the scanning electron microscopy (SEM) analysis seems more appropriate than the TEM investigation. Scattering techniques such as small angle neutron scattering (SANS) can be used to study large ensembles, but sensitivity of the scattering experiment to other elements introduces noise and uncertainty, and results are often prejudiced by the models used in analysis. Another procedure in grain size determination is peak shape analysis in the X-ray diffraction technique, but this is suitable only for particles with grain size less than 50 nm. In fact it is necessary that the coherent scattering corresponds to the grain size. That not being our case, X-ray diffraction technique has not been taken into account. Magnetometry is suitable for macroscopic measurements of a complete ensemble of particles, but this measurements interpretation and their correlation with the grain size distribution is less direct than that involved in the SEM and TEM approach.

The Brunauer-Emmett-Teller (BET) measurement seems the fastest, the cheapest and the most reliable technique to determine the specific surface of porous and very fine solid materials [14, 15].

Another fast technique could be the light scattering approach; unfortunately this methodology presents difficulty to disperse homogeneously the $MgB_2$ powders in a suitable solvent. In fact $MgB_2$ is easily decomposed reacting with water or oxidized by many of the common solvents, furthermore it gives aggregation phenomena.

All the techniques here reported are well known and typically used in separate way by other authors to investigate the morphology, the phase purity and the grain size: i.e. Hanson *et al* estimated particle sizes from field cooled/zero field cooled (FC/ZFC) experiments [16], and Lu *et al*. have suggested [17] one way to use these methods for the volume distribution evaluation. Häßler *et al* used the X-ray diffraction technique to study the $MgB_2$ grain size [18], Xu *et al* and Soltanian *et al* reported the grain size distribution calculated both from SEM and TEM image analysis [11, 19].

To summarize, in the present paper we investigate two common techniques for particles size determination, SEM and BET and we compare the data with results obtained by an innovative method based on the fitting of magnetization versus temperature curves, taking into account the Ginzburg-Landau temperature dependence of the London penetration depth. In order to determine the advantages and the disadvantages for each technique the corresponding data have been compared. The BET analysis has been also used to determine the boron precursor grain size.

## 2. Experimental details

*2.1. Boron precursors and $MgB_2$ synthesis*

The measurements were performed on two different powder typologies: standard laboratory synthesized $MgB_2$ powder (SP) and not standard laboratory synthesized $MgB_2$ powder (NSP). The SP $MgB_2$ sample was synthesized using commercial precursors (Mg and B). The not standard $MgB_2$ powders were prepared at different doping grade, pure (NSP0), C-doped 6% volume/volume (v/v) (NSP6) and C-doped 10% v/v (NSP10) respectively, reacting commercial Mg with the (C-doped) laboratory-made B synthesized by the new route described in [5]: briefly it consists in a nanostructuration step of the precursor boron oxide ($B_2O_3$) by spraying a water solution of (C-doped) $B_2O_3$ in liquid nitrogen and freeze-drying it before its reduction to nano-structured B by magnesiothermic reduction, well known as Moissan's process [20]. Moissan's process is the most important boron synthesis technique for large scale production of this element. As C source chitosan has been used. The C doping has a double purpose: to produce a homogeneous dispersion of nanometric C-aggregate in the B matrix and C substitution on B site. Both these kinds of defects have been produced to enhance the superconducting properties of the final $MgB_2$. However the superconducting characterization is not the main goal of this paper, and it was reported elsewhere [6].

Independently on the nature of the B, commercial or lab-made, the same parameters were adopted in the $MgB_2$ synthesis: 1 h at 920 °C under argon flow in a furnace directly connected to a glove box.

The $MgB_2$ powder samples have been summarized in table 1.

**Table 1.** List of the $MgB_2$ powder samples prepared and used for the grain size analysis.

| Powder samples | |
| --- | --- |
| SP | Standard Powder from commercial B (H. C. Starck grade I) |
| NSP0 | Not Standard Powder from pure laboratory made boron |
| NSP6 | Not Standard Powder from C-doped (6% v/v) laboratory made boron |
| NSP10 | Not Standard Powder from C-doped (10% v/v) laboratory made boron |

## 2.2. SQUID magnetometer measurements

A commercial 5.5 T MPMS Quantum Design Squid magnetometer was employed for the magnetization vs. temperature measurements at 1 mT and temperatures ranging from 5 to 50 K: small amount of $MgB_2$ powder (20 mg) was dispersed into the Apiezon® N grease using agate pestle and mortar in order to disaggregate the agglomerated particles into single grains. Each experimental curve of FC magnetization vs. temperature, representing the transition from the normal state to the superconducting state, has been fitted with the theoretical magnetization ($m(T)$) calculated taking into account two contributions $P(T)$ and $Q(T)$; which are correlated by the equation (1):

$$m(T) = Q(T) * P(T) \qquad (1)$$

in which $Q(T)$ is the superconducting diamagnetic response curve of the material, for our purpose it can be approximated by a step function (2).

$$Q(T) = \begin{cases} A & \text{if } T < T_C \\ 0 & \text{if } T \geq T_C \end{cases} \qquad (2)$$

$P(T)$ is the effective variation of the volume penetrated from the magnetic field (see figure 1), that can be represented by function of expression (3):

$$P(T) = \begin{cases} \left(\dfrac{d - 2\lambda(T)}{d}\right)^m & \text{if } \lambda < \dfrac{d}{2} \\ 0 & \text{if } \lambda \geq \dfrac{d}{2} \end{cases} \qquad (3)$$

Where $d$ is the average particle diameter $<d>$ and the penetration depth ($\lambda$) as function of temperature is represented by equation (4):

$$\lambda(T) = \dfrac{\lambda_0}{\sqrt{1 - \left(\dfrac{T}{T_C}\right)^n}} \qquad (4)$$

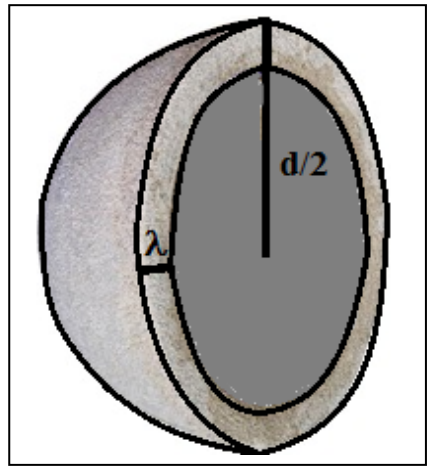

**Figure 1.** Cross section of a $MgB_2$ grain, the dark grey region represents the effective volume of the $MgB_2$ grain that contributes to the magnetization. Light grey region represents the volume penetrated by the magnetic field.

In our case the grains are three-dimensional then m = 3; but the method could be applied also in case of plate-like (e.g. Bi(2223) superconducting phase) or needle-like grains adopting m = 2 or 1 respectively. The power index $n$ into the analytical expression (4) represents different mechanisms: $n$ = 4 is for a strong coupling, $n$ = 3 is for a weak coupling, $n$ = 2 corresponds to an unconventional pairing state of a high level of impurities, in case of the MgB$_2$ the *n* value is still under debate. Using different measurement apparatus several groups reported *n* values ranging from n = 1 to 4 [21-24]. For the *m(T)* fit we used *n* = 1, 1.5 and 2, evaluating several $\lambda_0$ value: 85, 110, 180 nm found in the literature [23, 25-27].

*2.3. Scanning Electron Microscopy investigations and image analysis*

The microstructure of MgB$_2$ powders after the synthesis process was investigated using Leica Cambridge Scanning Electron Microscopy S 360 (SEM). The powders subjected to the SEM analysis have been pelletized in order to have a suitable planar surface for the grain counting and to evaluate the grain size we performed a sampling on about 8 SEM images taken from different areas of each sample. In particular the pictures have been acquired at different magnification: 5000x, 10000x and 20000x. The 10000x magnification was chosen as the most representative for the statistical image analysis. Then, the average grain diameter ($< d >$) of the powders has been determined counting the diameter of the grains in each picture both by a Matlab homemade routine and manually using the Image-Pro Express software.

The first step in the Matlab routine is to maximize the intensity contrast in the SEM acquired image and then rescale the image intensity so that it fills the data type's entire dynamic range. The second step is to determine the intensity surface area distribution in the enhanced image. The routine likens image objects to circular stones whose size can be determinated by sifting them through screens of increasing size and collecting what remains after each pass. Image objects are sifted by opening the image with a structuring element of increasing size and counting the remaining intensity surface area (summation of pixel values in the image) after each opening. We choose a counter limit so that the intensity surface area goes to zero increasing the size of structuring element. A significant drop in intensity surface area between two consecutive openings indicates that the image contains objects of comparable size to the smaller opening. This is equivalent to the first derivative of the intensity surface area array, which contains the size distribution of the MgB$_2$ grains in the image. The minima of the first derivative of the intensity surface area array tell us that MgB$_2$ grains in the image have those radii. The more negative the minimum point, the higher the MgB$_2$ grains cumulative intensity at that radius.

The particle diameters, estimated manually or automatically, have been statistically analyzed to give a frequency count value as function of the diameter, each distribution has been fitted with a Lorentzian curve, as reported in figures 6 and 8. In order to have a direct comparison and more information, frequency count values have been turned into volume percent vs. particle diameters, figures 7 and 9.

*2.4. Brunauer-Emmett-Teller measurements*

The Brunauer-Emmett-Teller (BET) method is the most widely used procedure for the determination of the surface area of solid materials and involves the use of the BET equation:

$$\frac{1}{W*\left(\frac{P_0}{P}-1\right)} = \frac{1}{W_m*C} + \frac{C-1}{W_m*C}*\left(\frac{P}{P_0}\right) \qquad (5)$$

in which *W* is the weight of adsorbed gas at a relative pressure *P/P$_0$*, and *W$_m$* is the weight of the adsorbate constituting a monolayer of surface coverage. The term *C* represents the BET *C* constant, which is related to the energy of adsorption in the first adsorbed layer and consequently its value is an indication of the magnitude of the adsorbent-adsorbate interactions. Before to perform the BET measurement each sample was subject to heat treatment in order to evacuate eventual adsorbed gasses, being the powders stored under argon atmosphere after their synthesis a short degassing time of about 30 min at 120 °C has been necessary.

For each method (SQUID, SEM and BET) it has been hypothesized that the particles have a spherical shape, in fact if the disk shape was considered as the real shape then the thickness of the MgB$_2$ disks should be determined, and unfortunately it is not easy to know this dimension. In the latter case the only one suitable technique would be the SEM image analysis.

### 3. Results

*3.1. SQUID magnetometer measurements*

Better results of the fitting were obtained using the exponential parameter $n = 1.5$ and $\lambda_0 = 85$ nm. The corresponding fit for the four powders are reported in figure 2; meanwhile the parameters extracted from the fit are summarized in table 2.

**Table 2.** Fit parameters from SQUID measurements on $MgB_2$ powders.

| Powder | n | $\lambda_0$ [nm] | $<d>$ [nm] | $T_C$ [K] | A [A/m] | err [A/m] |
|---|---|---|---|---|---|---|
| SP | 1.5 | 85 | 2770 | 37.9 | -1012 | 131 |
| NSP0 | 1.5 | 85 | 761 | 35.5 | -5611 | 877 |
| NSP6 | 1.5 | 85 | 795 | 32.1 | -3288 | 986 |
| NSP10 | 1.5 | 85 | 374 | 32.5 | -5129 | 329 |

The powders with the laboratory-made boron are finer than the $MgB_2$ with commercial B, which shows the sharpest transition curve. The difference in grain size in our powders, obtained using the same procedure for all samples, can be explained keeping into account that the C nano-aggregate in the liquid Mg can act as nucleation centre during the B synthesis. So the pure NSP0 has bigger grains than C-doped powders. In other words, the higher the C amount the finer the $MgB_2$ powder.

It is clear that the resolution of this method is the penetration depth (85 nm). Particles with a diameter less than 85 nm are fully penetrated by the magnetic field and thus cannot be recognized.

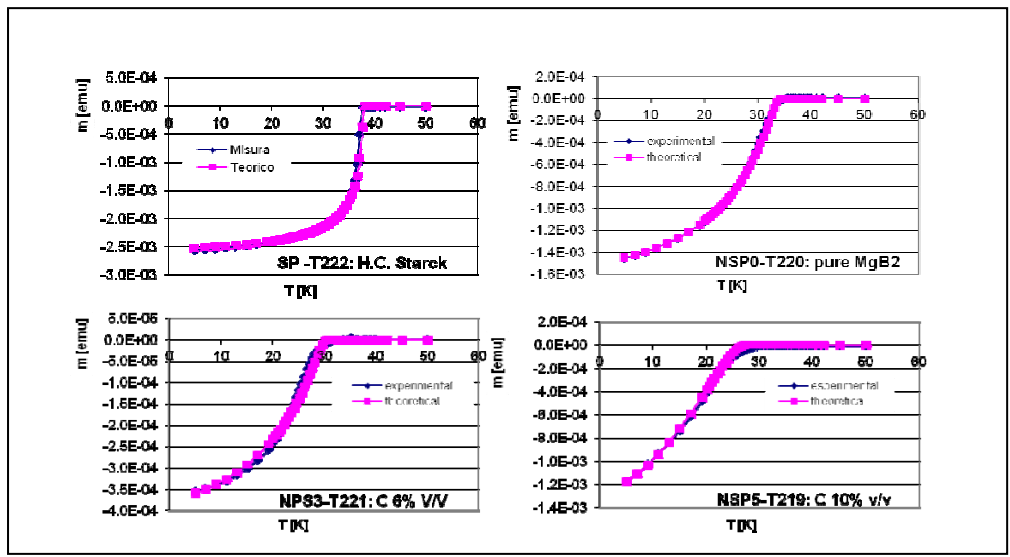

**Figure 2.** Comparison among the magnetization fit of different kind of powders: (a) SP, (b) NSP0, (c) NSP6 and (d) NSP10.

So, the average grain size of nanometric samples is overestimated in respect to those on micrometric scale. It is worth noting how the different NSP powders have more or less the same $T_C$ value. This behavior may be due to the nano-structuration of the lab-made B.

*3.2. SEM investigation and image analysis*
In order to have the widest view about the powder morphology in figure 3 the SEM image at 5000x is reported for each powder typology. The three $MgB_2$ powders synthesized using lab-made B appear finer than the $MgB_2$ obtained from the commercial B. In particular the pure NSP0 powder has a better shaped morphology than C-doped powders (NSP6 and NSP10); in figure 3(b) it is possible to see some hexagonal grains come out from the image's plane. The SP sample shows some aggregates, which are formed by particles with 0.5-2 μm in diameter.

In figure 4 the SEM images at 10000x are reported. The NSP10 sample is always the finest powder, in agreement with the magnetization response, even if some few big grains (1-2 μm) are present. Figure 5 shows the real nature of those big grains, which result to be agglomerates of very fine particles.

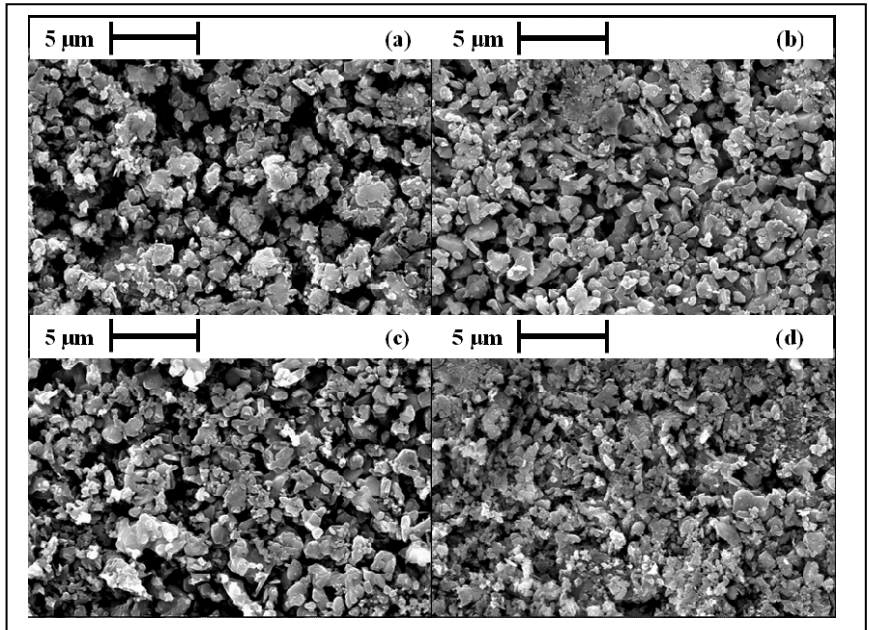

**Figure 3.** SEM images of the MgB$_2$ powders at 5000x: (a) SP, (b) NSP0, (c) NSP6 and (d) NSP10.

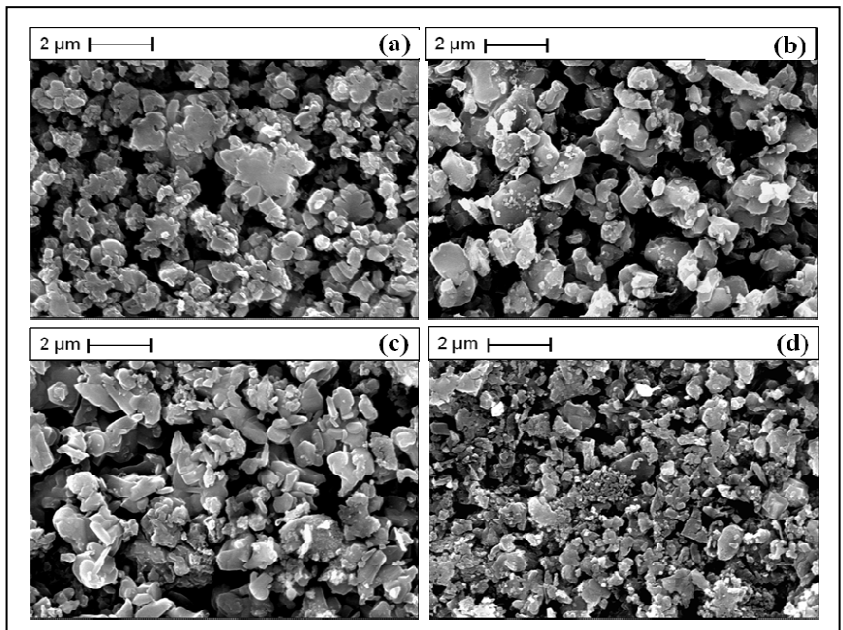

**Figure 4.** SEM images of the MgB$_2$ powders at 10000x: (a) SP, (b) NSP0, (c) NSP6 and (d) NSP10.

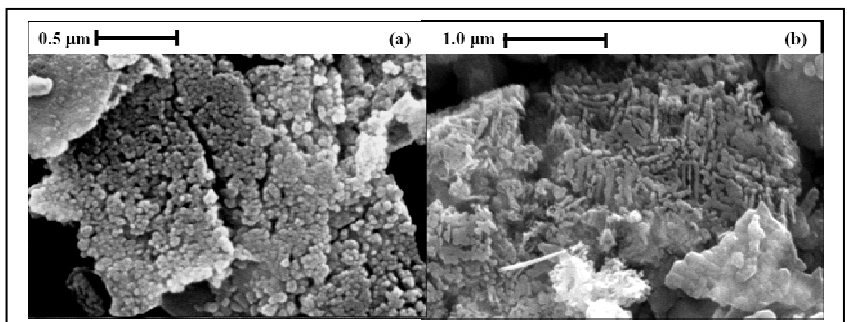

**Figure 5.** Particular at higher magnification of: (a) NSP6 at 50000x and (b) NSP10 at 30000x.

A manual statistical count of particle diameters has been performed on SEM images using the Image-Pro Express software. The 10000x magnification has been chosen as the most appropriate for both methods: manual and automatic count. The results are expressed in figure 6 as intensity *vs.* particle diameter. The data of the manual measurement are reported in table 3 as average grain diameter $< d >$. This value represents the centre of the Lorentzian fit ($x_o$). The related width of the peak (± FWHM) is also reported into the same column. The next column on the right of < $d$ > values reports the range of measured diameter, i.e. the variance. The variance is a measure of how far a set of numbers is spread out. The Lorentzian fit is also suitable to have a direct comparison between the different samples.

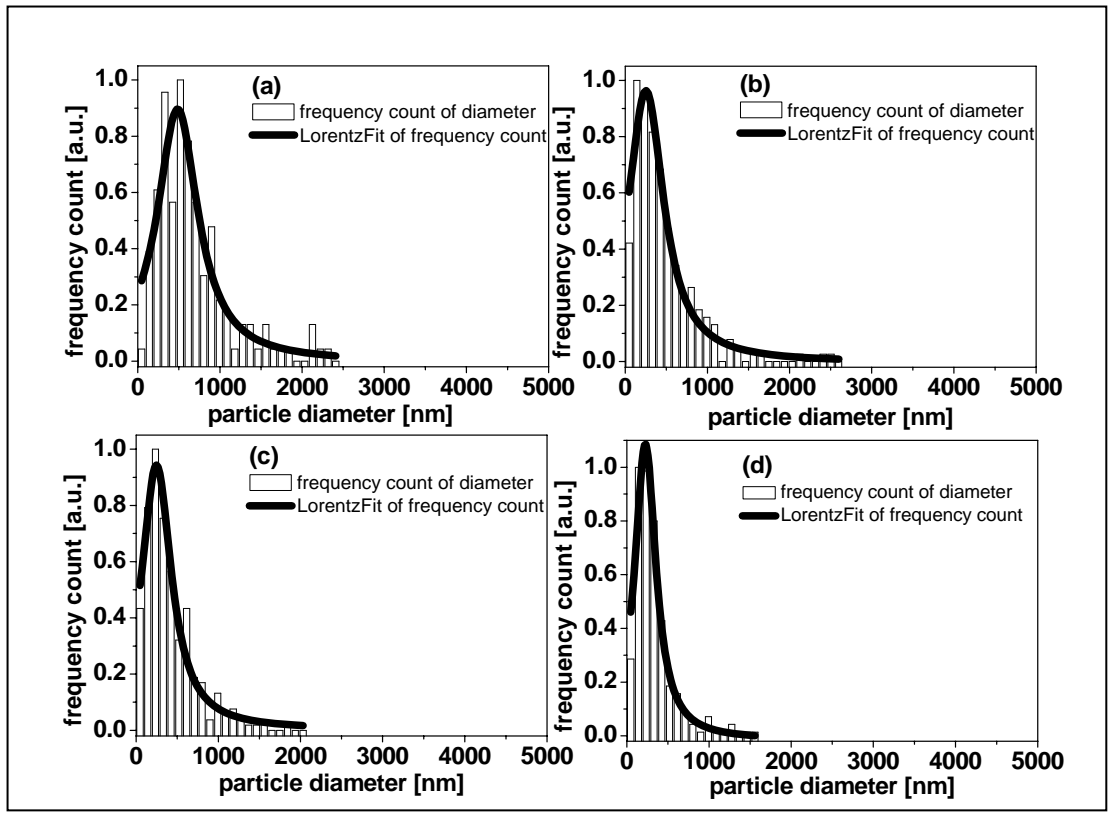

**Figure 6.** Grain size distribution (intensity vs. particle diameter) by manual measurement of the particle's diameter: (a) SP, (b) NSP0, (c) NSP6 and (d) NSP10.

In order to give more emphasis to the variance (to keep into account the presence of few big grains), the volume% has been calculated for each powder considering the ratio between the volume occupied by each particle having the same diameter and the volume occupied by all the particles counted. In this way each distribution is given as the sum of volume% *vs.* the particle diameter, see figure 7. It is noteworthy that data of figure 6 represent a distribution of the particles for each grain size; so, using this representation mode, the data interpretation could be ambiguous. In fact the volume occupied by each group of particles must be considered. The representation of the data in figure 7 looks to be the most complete method to represent the grain size. Figure 7 shows that the 100% of the NSP10 particles has a diameter less than 1500 nm, as well as the 60% of them is lower than 1000 nm and so on. For example the 100% of SP particles is under 2200 nm and the 60% of them is less than 2000 nm.

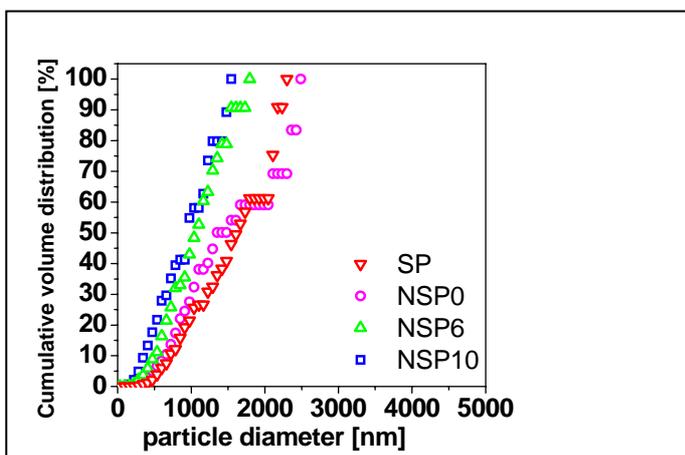

**Figure 7.** Comparison of volume% sum by manual measurements of the particle's diameter.

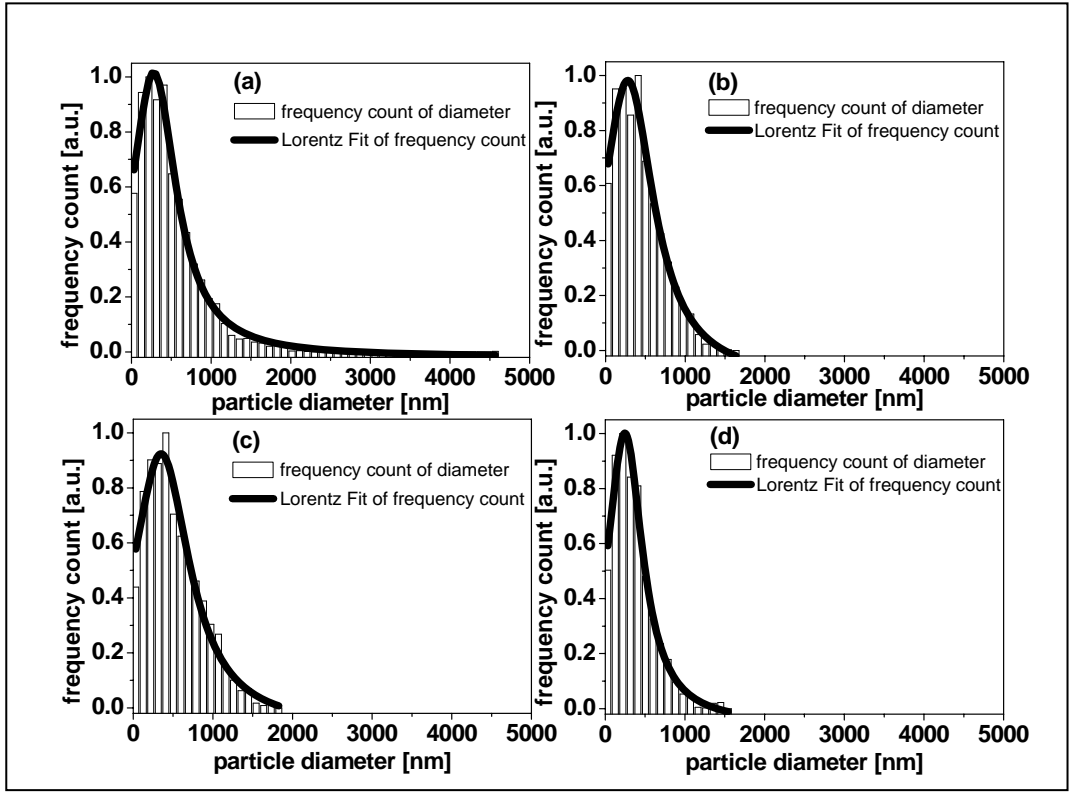

**Figure 8.** Grain size distribution (intensity vs. particle diameter) by automatic measurement of the particle's diameter: (a) SP, (b) NSP0, (c) NSP6 and (d) NSP10.

Figures 8 and 9 report the intensity as a function of particle diameter and the sum of volume% *vs.* the particle diameter respectively, obtained by automatic software count. It is interesting noting how the results by automatic count are more in agreement with the magnetic measurements than the manual ones. Of course manual count is operator dependent, and so it is really hard to measure all the particles within the image on a big amount of grains.

In figure 9 the 100% of each sample is under: 1500, 1600, 1900 and 4500 nm respectively for NSP10, NSP0, NSP6 and SP powder. SP sample has a continuous distribution of particles between 2000 and 4500 nm. The presence of these particles indicates that the SP powder is not a homogeneous sample and the 30% of the volume of this powder is occupied by grains with a diameter greater than 2000 nm.

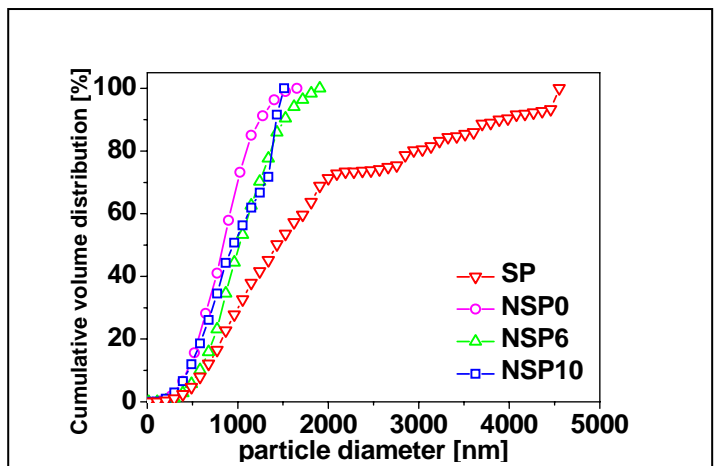

**Figure 9.** Comparison of volume% sum by automatic measurement of the particle's diameter.

3.3. *BET measurements*
The results of the BET measurements, expressed in $m^2/g$ were converted into the average diameter $<d>$ using the formula (6):

$$<d> = \frac{6}{\rho * S} \qquad (6)$$

In which $S$ is the BET specific surface area and $\rho$ is the true density, in the specific case the value of 2.63 g/cm$^3$ was used for the MgB$_2$ powders. The $<d>$ value of each sample obtained through the BET analysis is really similar; this is due to the spherical shape approximation of the MgB$_2$ grains. This technique is more suitable if applied on particles with homogeneous grain size. So, for agglomerated powders the $<d>$ value calculated by BET analysis will be underestimated. This is the case of SP sample. The data of the BET analysis have been reported in the last column of table 3 and 4. Probably when the powders are agglomerate with heterogeneous diameters the best technique is always the SEM image analysis.

The BET analysis was also employed to establish the grain size of B precursors. The commercial B results to have a $<d>$ of 200 nm, bigger than $<d>$ value of 60 nm of the pure lab-made B. A good agreement has been reached with SEM analysis results of reference [5]. The $<d>$ values of B have been calculated using $\rho = 2.45$ g/cm$^3$. It is worth to note that also in this case (as well as for the BET measurements on MgB$_2$) the real grain size of the commercial B may be higher than the calculated value; because also the commercial precursor has heterogeneous grain size distribution [5] and has the tendency to agglomerate. However it is interesting to note that the lab-made B is really on nanometric scale, but with the synthesis reaction of the MgB$_2$ the dimension of the final product has been increased (at the least) 7 times. In this case the synthesis reaction adopted is the standard method optimized in our laboratory on the commercial B. A suitable synthesis process must be developed yet on the lab-made B in order to preserve the finest grain size of the starting B.

## 4. Discussion

The results of the three techniques are summarized in table 3.

**Table 3.** Data resulting from different techniques.

| MgB$_2$ powder sample | SQUID $<d>$ [nm] $\lambda_0 = 85$ nm; $n = 1.5$ | SEM $<d>$ [nm] manual count | variance on $<d>$ [nm] manual | SEM $<d>$ [nm] automatic count | variance on $<d>$ [nm] automatic | BET $<d>$ [nm] |
|---|---|---|---|---|---|---|
| SP | 2770 | 500 ± 609 | 0 to 2500 | 300 ± 606 | 0 to 4500 | 500 |
| NSP0 | 750 | 250 ± 528 | 0 to 2600 | 350 ± 636 | 0 to 1600 | 650 |
| NSP6 | 800 | 250 ± 439 | 0 to 2000 | 350 ± 865 | 0 to 2500 | 600 |
| NSP10 | 350 | 200 ± 305 | 0 to 1500 | 250 ± 360 | 0 to 1600 | 400 |

Both manual and automatic SEM measurements of figure 6 and 8 are reported as frequency count *vs.* particles diameter. With this data representation the variance of the particles diameters recognized by the statistical count have to be considered, not only the average grain size value $<d>$ ± FWHM. In fact it is important to give the information that the $<d>$ of NSP10 is more or less 200 nm, but few big particles are also present, this is particularly true for the SP powder. Probably the most suitable method for a reader to receive the results of SEM analysis is that given in figures 7 and 9, and summarized in table 4.

Table 4 reports the SEM results as d50 (sum volume% *vs.* particles diameter), i.e. the diameter value corresponding to the half population that lies under this value. But is also important to highlight that a huge approximation is employed: i.e. we are considering the spherical shape as the real shape of the MgB$_2$ grains. So, looking at the analysis results it will be more correct to report data giving $<d>$ ± FWHM and its variance. In this way the error on the measurement will be less than the error corresponding to the volume% sum, which is calculated as $V \approx r^3$.

**Table 4.** Data resulting from different techniques.

| MgB$_2$ powder sample | SQUID $<d>$ [nm] $\lambda_0 = 85$ nm; $n = 1.5$ | SEM $d_{50}$ [nm] manual count by vol% sum | SEM $d_{50}$ [nm] automatic count by vol% sum | BET $<d>$ [nm] |
|---|---|---|---|---|
| SP    | 2770 | 1600 | 1450 | 500 |
| NSP0  | 750  | 1350 | 830  | 650 |
| NSP6  | 800  | 1050 | 1010 | 600 |
| NSP10 | 350  | 945  | 945  | 400 |

Independently from the applied technique, the finest powder is always the NSP10 sample with an average grain size included in the range of 200 up to 400 nm, see table 3 and 4. Powders NSP0 and NSP6 have more or less the same dimension in each of the three techniques. The SP sample has the biggest grains both for the SQUID technique and SEM investigation; instead for the BET analysis the SP powder has a grain size between the NSP10 and NSP0, NSP6. The disagreement respect to the other analysis of the BET examination on SP powder has been explained considering this sample not homogeneous in grain size for the presence of agglomerated. Table 5 reports the diameter corresponding at different volume% sum ($d_i$, $i$ = 90, 50 and 10%).

**Table 5.** Data resulting from statistical calculation on SEM images.

| MgB$_2$ powder | $d_{90}$ [nm] manual | $d_{50}$ [nm] manual | $d_{10}$ [nm] manual | $d_{90}$ [nm] automatic | $d_{50}$ [nm] automatic | $d_{10}$ [nm] automatic |
|---|---|---|---|---|---|---|
| SP    | 2170 | 1600 | 730 | 3980 | 1440 | 675 |
| NSP0  | 2400 | 1350 | 660 | 1250 | 830  | 455 |
| NSP6  | 1535 | 1050 | 472 | 1500 | 1010 | 590 |
| NSP10 | 1480 | 945  | 340 | 1400 | 945  | 455 |

With this representation of the SEM data, manual and automatic, the grain size trend results to be SP > NSP0 ≈ NSP6 > NSP10. Figure 10 reports the trend for each single technique: SQUID, SEM (manual and automatic count) and BET. It is interesting to note that the trend for each technique is the same SP > NSP0 ≈ NSP6 > NSP10, with the exception of the SP value resulting from BET analysis.

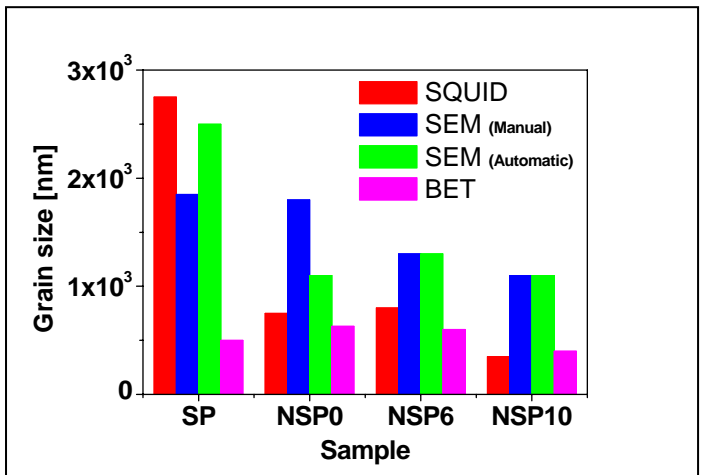

**Figure 10.** Comparison of $<d>$ value resulting from different techniques.

It will be indifferent to use the manual or the automatic counting method, of course the automatic methodology will be more suitable and operator independent.

In conclusion:

(i) The adopted technique will not be influent on the grain size determination in samples having homogeneous particles.
(ii) SEM imaging and SQUID measurements will be the best techniques for samples with not homogeneous particles.

## 5. Conclusions

In this paper, we investigate two common techniques for particle size determination: SEM investigation and BET analysis. Then we compare their results with those obtained by a less common method: magnetic measurements by SQUID in order to determine for each technique the advantages and the disadvantages.

The dependence of the human operator has been overcome by developing a Matlab routine to count automatically the intensity of the particles diameter of each powder in every acquired picture.
The most suitable way to establish the average grain size seems to be the contemporary use of the SEM image analysis (better by automatic count) and the SQUID magnetometer technique, in order to have a validation. The detectable limit in SQUID measurements is represented by the $\lambda_0$ value.

The best option for short time and low cost measurement would be to use the BET technique, but we saw that it is not a suitable choice for heterogeneous powders.

The driving force of this article is the important role played by the grain size. In consequence of this, its knowledge appears to be fundamental in order to predict materials behavior. In particular for the $MgB_2$ is important to investigate the influence of the grain dimension on the pinning mechanism dominated by the grain boundary pinning model [7, 28].

**Acknowledgments**

This work was supported in part by grants from Fondazione CARIGE, CNR-SPIN, University of Genova, and PRIN 2008.